\documentclass[epj,nopacs]{svjour}

\usepackage{amsmath, amssymb} 
\usepackage{physics} 
\usepackage{xcolor} 
\usepackage{graphicx}
\usepackage{tikz}
\usetikzlibrary{decorations.markings}

\begin{document}
\raggedbottom

\title{Constraints on Yukawa-type New Forces from the Lamb Shift in Muonic Hydrogen and Deuterium}
\author{Chung-Chien Huang\inst{1} \thanks{\emph{email:} bingo.aplus@gmail.com} \and Li-Bang Wang\inst{2} \thanks{\emph{email:} lbwang@phys.nthu.edu.tw}
}     

\institute{National Experimental High School at Hsinchu Science Park, Hsinchu, 300094, Taiwan \and Department of Physics, National Tsing Hua University, Hsinchu, 300044, Taiwan}

\date{Received: date / Revised version: date}

\abstract{
We develop an analytical framework to constrain Yukawa-type fifth-force interactions using the Lamb shift in hydrogen and deuterium. The sixfold coordinate-space integral for the energy shift is transformed into a one-dimensional momentum-space convolution of the atomic form factor and the Yukawa potential, eliminating the need for approximations. This formulation directly incorporates empirical nuclear charge distributions through their form factors, making the implementation of experimental results straightforward. Applying this method to the latest hydrogen and deuterium spectroscopy data, we derive constraints on the coupling constant $\alpha$ in the interaction range $\lambda \in [10^{-4}, 10^{6}]$~fm. A resonance-like cancellation between the $2s$ and $2p$ orbital shifts is identified at a critical range $\lambda \sim 10^{5}$~fm, where the energy shift vanishes and the constraint on $\alpha$ diverges, marking a transition between attractive and repulsive regimes. The constraints are shown to be independent of the nuclear charge distribution model. Our results indicate that the transition frequency shifts induced by such new forces lie well below current spectroscopic resolution, establishing spectroscopic precision as the primary bottleneck for probing these interactions.
}

\titlerunning{Constraints on Yukawa-type New Forces}
\maketitle

\section{Introduction}
The discrepancy between the proton charge radius of normal and muonic hydrogen measured via spectroscopy in the early 2010s, widely recognized as the "Proton Radius Puzzle", has motivated investigations into physics beyond the Standard Model. Among the most outstanding candidates are modified gravity theories \cite{ME,Wang} and the introduction of a "fifth force." \cite{1,2,3,4,5,6} Such interactions are frequently modeled as a Yukawa-type potential, typically arising from the exchange of light scalar or vector bosons with mass. While efforts to constrain these interactions have been made by many experiments, including torsion balance tests, atomic spectroscopy, and astrophysical observations, the parameter space for the coupling strength $\alpha$ and the interaction range $\lambda$ remains largely unconstrained, particularly in the regime of short-range interactions where the Yukawa potential diverges at the origin.

In this paper, we managed to map the discrepancy of the measured proton and deuteron radii to the energy shift of Lamb shift introduced by the Yukawa-type potential. While the physical motivation is clear, the practical application of these potentials to atomic systems involves significant computational complexity which requires a six-fold spatial integral. Previous investigations \cite{roberto1,roberto2} often rely on approximations in coordinate space, such as assuming sphere-like nuclear charge distributions or truncation of the potential to estimate the energy shifts in the $2s$ and $2p$ states. These approximations inherently limit the precision of the resulting bounds on the coupling strength $\alpha$ and the interaction range $\lambda$, as they fail to capture the physical behavior of the Yukawa potential.

To overcome these limitations, we formulate the energy shift as a convolution of the atomic wave functions and the interaction potential. By transforming this interaction into momentum space, the convolution integral simplifies into a direct one-dimensional integral of the product of form factors. This approach allows for a precise, analytical solution for the Lamb shift discrepancy, effectively mapping the interaction's impact on atomic energy levels without the loss of precision associated with coordinate-space approximations. While the $2p$ orbital is traditionally assumed to be unaffected by the Yukawa potential due to its vanishing probability density at the origin, we identify a resonance-like cancellation effect between the energy shifts of the $2s$ and $2p$ states at a critical Yukawa range $\lambda$. This phenomenon induces a divergence in the coupling constant and marks the transition between attractive and repulsive interaction regimes. We performed a numerical scan over the Yukawa range $\lambda \in [10^{-4}, 10^{6}]$ fm for various nuclear distribution models, including the Gaussian and Dipole models. The results demonstrate that the constraints derived are highly independent of the nuclear distribution model, particularly in the low-momentum limit where the form factor converges to a universal behavior. This model independence ensures that our constraints on the Yukawa parameters are robust and reliable, providing a stringent test for new physics scenarios that may manifest as Yukawa-type interactions.

The remainder of this paper is organized as follows: In Sec. II, we detail the theoretical framework, in which the sixfold integral is transformed into a one-dimensional expression in momentum space. In Sec. III, we present the numerical constraints derived from the latest hydrogen and deuterium spectroscopy data, comparing the calculated Lamb shifts with current experimental resolutions.
\section{Theoretical Framework}
    \subsection{Real-to-Momentum Space Transformation}
        We consider a Yukawa-type interaction potential introduced by a mass distribution $\rho$, which can be expressed as :
        \begin{equation}
            V(r) = \alpha \int \rho(\vec{r}') \frac{e^{-|\vec{r} - \vec{r}'|/\lambda}}{|\vec{r} - \vec{r}'|} \dd^3 r'
        \end{equation}
        This potential provides an extra binding for the $2s$ and $2p$ orbitals. Within the framework of first-order perturbation theory, the required energy discrepancy is given by :
        \begin{equation}
            \Delta E_{\text{Lamb}} = \int (\rho_{2p}(\vec{r})-\rho_{2s}(\vec{r})) V(\vec{r}) \dd^3 r
        \end{equation}
        where $\rho_{2s}(\vec{r})$ and $\rho_{2p}(\vec{r})$ are the atomic charge densities of the $2s$ and $2p$ orbitals. Evaluating this integral is practically challenging due to the presence of a six-fold integration. To simplify the calculation, we transform the integral into momentum space. First, we express the atomic density and potential in terms of their Fourier transforms :
        \begin{align}
            \rho_{at} &= \frac{1}{(2\pi)^3} \int \tilde{\rho}_{at}(q) e^{i \vec{q} \cdot \vec{r}} \dd^3 q \\
            V &= \frac{1}{(2\pi)^3} \int \tilde{V}(q') e^{i \vec{q'} \cdot \vec{r}} \dd^3 q'
        \end{align}
        where $\rho_{at}$ is the atomic density for an arbitrary orbital. The energy shift can then be expressed as :
        \begin{align}
            \Delta E
            =& \frac{1}{(2\pi)^6} \int \int \int \tilde{\rho}_{at}(q) \tilde{V}(q') e^{i(\vec{q}+\vec{q'}) \cdot \vec{r}} \dd^3 q \dd^3 q' \dd^3 r \nonumber \\
            =& \frac{1}{(2\pi)^3} \int \int \tilde{\rho}_{at}(q) \tilde{V}(q') \delta^{3}(\vec{q} + \vec{q'}) \dd^3 q \dd^3 q' \nonumber \\
            =& \frac{1}{(2\pi)^3} \int \tilde{\rho}_{at}(\vec{q}) \tilde{V}(\vec{-q}) \dd^3 q \nonumber \\
            =& \frac{1}{(2\pi)^3} \int \tilde{\rho}_{at}(\vec{q}) \tilde{V}(\vec{q}) \dd^3 q
        \end{align}
        where we have implemented the symmetry of the potential $\tilde{V}(\vec{q}) = \tilde{V}(-\vec{q})$.

    \subsection{Fourier Transform of the Atomic Density Distributions}
        To accurately model the Lamb shift transition, we consider both the $2s$ and $2p$ orbitals. The atomic density distribution of the $2s$ and $2p$ orbitals is given by :
        \begin{align}
            \rho_{2s}(r) &= m_{lep} |\psi_{2s}(r)|^2 = m_{lep} \frac{1}{8\pi a_0^3} e^{-r/a_0} \left(1 - \frac{r}{2a_0}\right)^2 \\
            \rho_{2p}(r) &= m_{lep} |\psi_{2p}(r)|^2 = m_{lep} \frac{1}{96 \pi a_0^5} r^2 e^{-r/a_0}
        \end{align}
        where $a_0$ is the Bohr radius. It is important to note that while the $2p$ orbital wavefunction is not spherically symmetric, the Yukawa potential is purely radial, depending only on the radial distance $r$. In the calculation of the first-order energy shift, the probability density $|\psi_{2p}(r)|^2$ can be factorized into radial and angular components. Since the angular components are normalized, the integration over the solid angle yields unity. Consequently, the energy shift is determined solely by the radial component of the wavefunction, allowing the three dimensional wavefunction to be reduced to a one-dimensional radial distribution. The Fourier transform of an atomic density distribution $\rho_{at}(\vec{r})$ is given by :
        \begin{align}
            \tilde{\rho}_{at}(\vec{q}) &= \int \rho_{at}(r) e^{-i \vec{q} \cdot \vec{r}} \dd^3 r \nonumber \\
            &= \int_{0}^{\infty} \dd r \int_{0}^{\pi} \dd\theta \int_{0}^{2\pi} \dd\phi \, r^2 \sin\theta \, \rho_{at}(r) e^{-i q r \cos\theta} \nonumber \\
            &= \frac{4\pi}{q} \int_{0}^{\infty} r \sin(qr) \rho_{at}(r) \dd r
        \end{align}
        The Fourier transform $\tilde{\rho}_{2s}(\vec{q})$ and $\tilde{\rho}_{2p}(\vec{q})$ in the momentum space are expressed as :
        \begin{align}
            &\tilde{\rho}_{2s}(\vec{q}) = m_{lep} \frac{[2(q/b)^2 - 1][(q/b)^2 - 1]}{(1 + (q/b)^2)^4}  \\
            &\tilde{\rho}_{2p}(\vec{q}) = m_{lep} \frac{1 - (q/b)^2}{(1 + (q/b)^2)^4}
        \end{align}
        where $b = 1/a_0$. The difference in the Fourier transforms of the $2s$ and $2p$ orbitals is given by :
        \begin{equation}
            \tilde{\rho}_{2p}(\vec{q}) - \tilde{\rho}_{2s}(\vec{q}) = -m_{lep} \frac{2(q/b)^2[(q/b)^2 - 1]}{(1 + (q/b)^2)^4}
        \end{equation}

    \subsection{Fourier Transform of the Yukawa Potential and Nuclear Form Factors}
        The interaction also depends on the internal structure of the nucleus. We model the proton and deuteron charge distributions using two different models :
        \begin{itemize}
            \item \textbf{Gaussian Model :} $\rho_p(r) = m_p \left(\frac{3}{2\pi r_p^2}\right)^{3/2} e^{-\frac{3r^2}{2r_p^2}}$.
            \item \textbf{Dipole Model :} $\rho_p(r) = m_p \frac{3^{\frac{3}{2}}}{\pi r_p^3} e^{-2\sqrt{3} \frac{r}{r_p}}$.
        \end{itemize}
        Note that the normalization factors and the exponent coefficients are chosen to ensure that the charge distributions are properly normalized to the proton mass $m_p$ and the root-mean-square radius $r_p$. The Fourier transform of the potential function is obtained by convolving the Yukawa potential with the nuclear charge distribution :
        \begin{align}
            \tilde{V}(\vec{q}) 
            &= \int V(\vec{r}) e^{-i \vec{q} \cdot \vec{r}} \dd^3 r \nonumber \\
            &= \int \left[\alpha \int \rho_p(\vec{r}') \frac{e^{-|\vec{r} - \vec{r}'|/\lambda}}{|\vec{r} - \vec{r}'|} \dd^3 r' \right] e^{-i \vec{q} \cdot \vec{r}} \dd^3 r
        \end{align}
        We perform a variable substitution $\vec{R} = \vec{r} - \vec{r}'$ to simplify the expression :
        \begin{align}
            \tilde{V}(\vec{q}) 
            &= \int \left[ \alpha \int \rho_p(\vec{r}') \frac{e^{-R/\lambda}}{R} \dd^3 r' \right] e^{-i \vec{q} \cdot (\vec{R} + \vec{r}')} \dd^3 R \nonumber \\
            &= \left[ \int \rho_p(\vec{r}') e^{-i \vec{q} \cdot \vec{r}'} \dd^3 r' \right] \left[\alpha \int \frac{e^{-R/\lambda}}{R} e^{-i \vec{q} \cdot \vec{R}} \dd^3 R \right]
        \end{align}
        The first integral corresponds to the Fourier transform of the nuclear charge distribution, i.e. the product of the form factor and the mass of the nucleus. The second integral is the Fourier transform of the Yukawa potential. The Fourier transform of the Yukawa potential is given by :
        \begin{equation}
            \alpha \int \frac{e^{-R/\lambda}}{R} e^{-i \vec{q} \cdot \vec{R}} \dd^3 R = \frac{4\pi \alpha}{q^2 + 1/\lambda^2}
        \end{equation}
        The form factors for the Gaussian and Dipole models are given by :
        \begin{itemize}
            \item \textbf{Gaussian Model :} $F(q) = e^{-\frac{q^2 r_p^2}{6}}$.
            \item \textbf{Dipole Model :} $F(q) = \left(1 + \frac{q^2 r_p^2}{12}\right)^{-2}$.
        \end{itemize}
        It is noteworthy that in the low-momentum limit ($q \to 0$), the form factor of any arbitrary mass distribution approaches a universal behavior, becoming independent of the internal structure of the nucleus. In this limit, the form factor can be approximated as $F(q) \approx 1 - \frac{q^2 r_p^2}{6}$, which is consistent with the leading-order expansion of both the Gaussian and Dipole models. Such consistency in the low-momentum limit ensures that our results are robust and insensitive to the choice of the nuclear charge distribution model, particularly for long-range interactions where the momentum transfer is small.

    \subsection{Energy Shift Calculation}
        The energy shift $\Delta E$ can be expressed as :
        \begin{equation}
            \Delta E_{\text{Lamb}} = \frac{1}{(2\pi)^3} \int \left[\tilde{\rho}_{2p}(\vec{q}) - \tilde{\rho}_{2s}(\vec{q})\right] \tilde{V}(\vec{q}) \dd^3 q
        \end{equation}
        Substituting the expressions for $\tilde{\rho}_{2p}(\vec{q}) - \tilde{\rho}_{2s}(\vec{q})$ and $\tilde{V}(\vec{q})$, we have :
        \begin{align}
            \Delta E_{\text{Lamb}} = - \frac{2 \alpha m_{lep} m_{nuc}}{\pi} \int_0^\infty &\frac{2(q/b)^2[(q/b)^2 - 1]}{(1 + (q/b)^2)^4} \nonumber \\& F(q) \frac{1}{q^2 + 1/\lambda^2} q^2 \dd q 
        \end{align}
        where $F(q)$ is the form factor corresponding to the chosen nuclear charge distribution model. The integral can be evaluated numerically for different values of the Yukawa range $\lambda$ and the coupling constant $\alpha$. We can evaluate the difference of the integral between normal and muonic atoms to obtain the energy shift discrepancy $\Delta(\Delta E_{\text{Lamb}})$. By comparing this theoretical energy shift discrepancy with the experimental discrepancy, we can set constraints on the Yukawa parameters $\alpha$ and $\lambda$.

    \subsection{Energy Shift Analysis for Hydrogen}
        The values of the proton radius used are as follow: 
        \begin{align}
            &\text{Normal Hydrogen : }0.833(10) \text{ (fm)}\\ 
            &\text{Muonic Hydrogen : }0.84087(39) \text{ (fm)}
        \end{align}
        The values of the proton radius are taken from the 2019 Lamb shift measurement \cite{NH} and the 2013 muonic hydrogen measurement \cite{MH}. The difference between these values is then used to calculate the corresponding energy shift $\Delta(\Delta E_{\text{Lamb}})$, which is compared with the theoretical energy shift derived from the Yukawa potential to set constraints on the coupling constant $\alpha$ and the Yukawa range $\lambda$. The limit is determined by ensuring that the theoretical energy shift does not exceed the experimental discrepancy, thus providing a limit on the strength of any potential new force mediated by a Yukawa-type interaction. 

        To derive the restricted domain of the parameters of the Yukawa-type potential, we utilize the energy shift formulas for the Lamb shift. For hydrogen, the $2S_{1/2, F=1}$ to $2P_{3/2, F=2}$ energy difference, $\Delta E_H$, is given by \cite{E_H} :
        \begin{equation}
            \Delta E_{\text{Lamb}}^H = 209.9779(49) - 5.2262 \cdot r^2_{p} + 0.0347 \cdot r^3_{p} \text{ meV}
        \end{equation}
        where $r_p$ is given in femtometers. By substituting the proton radius values into this formula, we can compute the residual energy shift $\Delta(\Delta E_{\text{Lamb}})$ :
        \begin{equation}
            \Delta(\Delta E_{\text{Lamb}}^H) = -0.069 \pm 0.087 \text{ meV}
        \end{equation}
        We take the upper bound $\Delta(\Delta E_{\text{Lamb}}^H) = 0.018$\,meV
        (one standard deviation) to derive the maximum allowed coupling
        constant $\alpha$ for a given Yukawa range $\lambda$.

    \subsection{Energy Shift Analysis for Deuterium}
        Since a universal Yukawa-type interaction would affect both hydrogen and deuterium, we also analyze the energy shift for Deuterium. As a discrepancy of the deuteron radius measurements is observed between the 2016 and 2017 spectroscopy measurements, we derive the required coupling constant $\alpha$ and the Yukawa range $\lambda$ that would contribute to the residual discrepancy. To determine whether the Yukawa-type interaction can account for the observed discrepancy, we examine the consistency between the restricted domain derived by the hydrogen system and the parameters required to explain the Deuterium discrepancy. The radius of the deuteron is as follows :
        \begin{align}
            &\text{Normal Deuterium : }2.1415(45) \text{ (fm)}\\ 
            &\text{Muonic Deuterium : }2.12562(78) \text{ (fm)}
        \end{align}
        The values of the deuteron radius are based on the 2016 and 2017 spectroscopy measurements \cite{ND,MD}. In order to map the discrepancy of the deuteron radius, we apply the energy shift formulas for the Lamb shift. For deuterium, the $2S_{1/2, F=1}$ to $2P_{3/2, F=2}$ energy difference, $\Delta E_D$, is given by \cite{MD} :
        \begin{equation}
            \Delta E_{\text{Lamb}}^D = 228.7766(10) + 1.7096(200) - 6.1103(3) \cdot r^2_{d} \text{ meV}
        \end{equation}
        where $r_d$ is given in femtometers. By substituting the deuteron radius values into this formula, we can compute the residual energy shift $\Delta(\Delta E_{\text{Lamb}})$ : 
        \begin{equation}
            \Delta(\Delta E_{\text{Lamb}}^D) = 0.42 \pm 0.12 \text{ meV}
        \end{equation}
        We take the upper bound $\Delta(\Delta E_{\text{Lamb}}^D) = 0.54$\,meV
        (one standard deviation) to derive the maximum allowed coupling
        constant $\alpha$ for a given Yukawa range $\lambda$.
\section{Results and Discussion}
    \subsection{Constraint Curves}
        \begin{figure}
            \centering
            \includegraphics[width=0.9\linewidth]{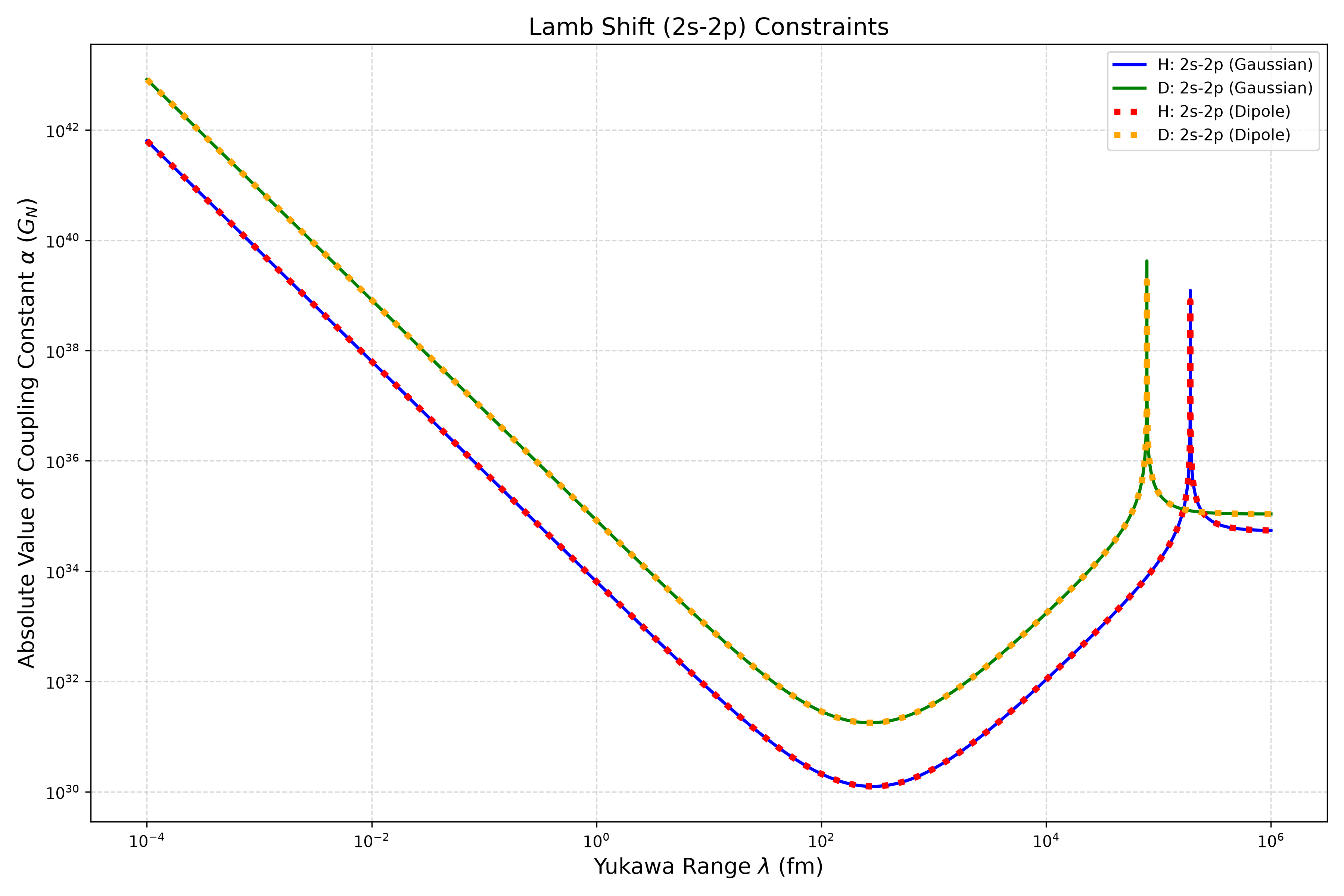}
            \caption{Yukawa coupling constant $\alpha$ as a function of the range $\lambda$. The region below the curves is the allowed parameter domain. The coupling constant $\alpha$ at the resonance point should be infinity. However due to limited resolution of the plot, only a finite value is shown.}
            \label{fig:constraint_dip}
        \end{figure}
        We performed a numerical scan over the Yukawa range $\lambda \in [10^{-4}, 10^{6}]$ fm. The maximum allowed coupling constant $\alpha$ is derived by comparing the theoretical shift with the radius of the normal and muonic hydrogen. By comparing the allowed parameter spaces, a significant discrepancy between the two atomic systems is observed.

        The limits on the Yukawa coupling constant $\alpha$ imposed by the $2S_{1/2}-2P_{3/2}$ Lamb shift measurements in hydrogen and deuterium for gaussian and dipole nuclear form factors are presented in Fig.~\ref{fig:constraint_dip}. 

        We observe that the constraints derived from the hydrogen system are more stringent than those from the deuterium system. This is primarily due to the larger discrepancy between the deuteron radius measurements compared to the proton radius measurements. The deuterium system exhibits a larger difference between the muonic and normal electronic atomic Lamb shifts, leading to a wider allowed parameter space for the coupling constant $\alpha$. In contrast, the hydrogen system has a smaller discrepancy in the radius measurements, resulting in tighter constraints on $\alpha$.
    \subsection{Transition Frequency Between the 2s and 2p Orbitals in Hydrogen}
        \begin{figure}
            \centering
            \includegraphics[width=0.9\linewidth]{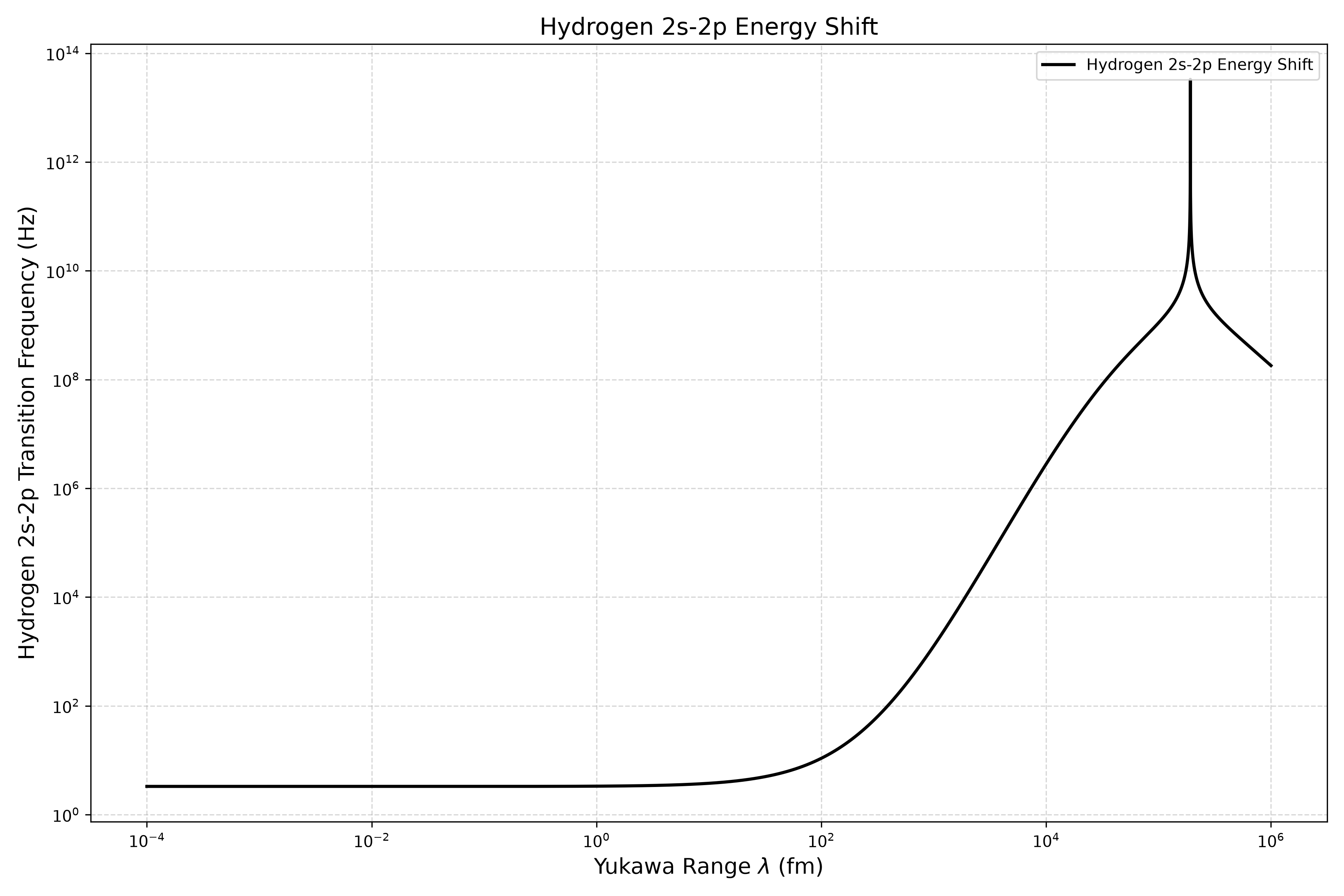}
            \caption{The transition frequency between the $2s$ and $2p$ orbitals in hydrogen as a function of the Yukawa range $\lambda$. The curve represents the prediction of the transition frequency shift that corresponds to the energy shift introduced by the Yukawa potential.}
            \label{fig:transition_frequency}
        \end{figure}
        We provide a plot of the transition frequency between the $2s$ and $2p$ orbitals in hydrogen as a function of the Yukawa range $\lambda$. The curve represents the prediction of the transition frequency change that corresponds to the energy shift introduced by the Yukawa potential. In the short-range limit ($\lambda \ll a_0$), i.e. the regime of physical interest, the transition frequency is around the magnitude of $10^1$ Hz. In such a regime, the energy shift of $2s-2p$ transition can be approximated to be equivalent to the energy shift of the $2s-np$ (np represents an arbitrary p orbital) transition, as the $2p$ orbital is not significantly affected by the Yukawa potential under short-range limit. Due to the small magnitude of the energy shift compared with the resolution of transition frequency measurements, which is around several kHz, the shift of transition frequency is expected to be extremely challenging to measure. 
    \subsection{Model Independence and Nuclear Form Factor Effects}
        To assess the robustness of our constraints against nuclear structure uncertainties, we compared the results derived from Gaussian and Dipole nuclear charge distributions. As in Fig.~\ref{fig:constraint_dip}, the constraints from different form factors converge perfectly. This demonstrates that our constraints are strictly model-independent for long-range modified gravity or fifth-force models. Such results are expected as we can see how the form factor converges to $F(q) \approx 1 - \frac{q^2 r_p^2}{6}$ under low-momentum limit. In other words, despite the nuclear structure model applied, the constraint derived will not vary tremendously.
    \subsection{Resonance-like Cancellation}
        A prominent and physically profound feature observed in the constraint plot is the emergence of a sharp resonance-like peak where the upper limits on the coupling constant $\alpha$ diverge to infinity. For hydrogen, the resonance is observed at $\lambda \approx 1.9 \times 10^{5} \text{ fm}$. While for deuterium, it occurs at $\lambda \approx 7.9 \times 10^{4} \text{ fm}$. We must emphasize that this is not a physical resonance associated with a particle state. Instead, it originates from the cancellation between the shift that the Yukawa-type potential introduces to $2s$ and $2p$ orbitals. We constrain the Yukawa parameters through the discrepancy between the muonic and normal electronic atomic Lamb shifts, setting the constraint condition as $\Delta(\Delta E_{\text{Lamb}}) = \Delta E^{\mu}_{\text{Lamb}} - \Delta E^{e}_{\text{Lamb}}$. When the difference of the energy shift of the Lamb shift approaches zero, resonance occurs. Beyond the resonance point, the coupling constant is expected to be positive, indicating that the Yukawa-type potential contributes to a repulsive interaction. Below the resonance point, the coupling constant is expected to be negative, indicating that the Yukawa-type potential contributes to an attractive interaction. Around the resonance point, no constraint can be set on the coupling constant $\alpha$, as the Yukawa-type potential does not contribute to the energy shift of the Lamb shift.

        While the cancellation intrinsically involves both systems, a rigorous scaling analysis reveals that the phenomenon is dominated by the muonic atom. The prefactor of the perturbative energy shift scales heavily with the lepton mass as $m_{lep}/a_0^3 \propto m_{lep}^4$. Meanwhile, the momentum integral, in the long-range limit ($\lambda \gg a_0$, as the resonance is observed at a Yukawa range that's several magnitude larger than the Bohr radius of muonic and normal atoms), is proportional to $\beta^2 \propto 1/m_{lep}^2$. The net energy shift thus scales approximately as $m_{lep}^2$. Consequently, for a given large Yukawa range, the perturbation in muonic hydrogen is fundamentally amplified by a factor of $(m_\mu/m_e)^2 \approx 4.3 \times 10^4$ relative to normal hydrogen ($\Delta E_{\mu} \gg \Delta E_{e}$). The exact macroscopic cancellation condition $\Delta E_{\mu} = \Delta E_{e}$ can be said to be equivalent to the self-cancellation of the muonic energy shift alone : $\Delta E_{\mu} = \Delta E_{2s,\mu} - \Delta E_{2p,\mu} \approx 0$.

        To rigorously understand this phenomenon, we examine the momentum-space integral that governs the energy shift. Using dimensionless momentum $x = q a_0$ and the interaction range parameter $\beta = a_0/\lambda$, the normalized energy shift takes the functional form :
        \begin{equation}
        \Delta E_{\text{Lamb}} \propto \int_0^\infty \frac{x^4(x^2 - 1)}{(x^2+1)^4 (x^2 + \beta^2)} F(x) \, dx \equiv I(\beta)
        \end{equation}
        We derive an approximated scaling analysis under the assumption of a point mass distribution. In the limit of a point-like nucleus where the form factor is trivial ($F(x) = 1$), we can evaluate the base integral $I(\beta)$ analytically. Because the integrand is an even function, the integral from $0$ to $\infty$ is half the integral over the entire real axis. Extending to the complex plane and applying the Cauchy Residue Theorem over the upper half-plane, we sum the residues evaluated at the simple pole $z=i\beta$ and the fourth-order pole $z=i$. The detailed derivation is provided in \ref{app:analytical_derivation}. After algebraic simplification, the integration yields a strictly positive, exact analytical form :
        \begin{equation}
        I_{\text{point}}(\beta) = \int_0^\infty \frac{x^4(x^2 - 1)}{(x^2+1)^4 (x^2 + \beta^2)} \, dx = \frac{\pi \beta^2}{8(\beta+1)^4}
        \end{equation}
        Since the scale parameter $\beta = a_0/\lambda > 0$, $I(\beta)$ is rigorously positive. Physically, the $2s$ state possesses a non-zero probability density at the origin , whereas the $2p$ orbital vanishes at the origin. Thus, the $2s$ state consistently experiences a larger attractive Yukawa potential well than the $2p$ state. In the point-like nucleus approximation, this energy difference is permanently positive; the constraint curve would not experience any resonant features.

        The existence of the peak is caused entirely by the finite volume of the nucleus. For a typical Gaussian nuclear form factor (or any arbitrary non-point-like distribution), its low-momentum expansion is given by $F(x) \approx 1 - \gamma x^2$, where the dimensionless finite-size parameter is defined as $\gamma = \frac{1}{6}\left(\frac{r_{nuc}}{a_0}\right)^2$. Substituting this expansion into the total integral yields a first-order suppression :
        \begin{align}
        I(\beta) &\approx \int_0^\infty \frac{x^4(x^2 - 1)}{(x^2+1)^4 (x^2 + \beta^2)} (1 - \gamma x^2) \, dx \\
        &= I_{\text{point}}(\beta) - \gamma J(\beta)
        \end{align}
        where $J(\beta)$ represents the integral :
        \begin{equation}
        J(\beta) = \int_0^\infty \frac{x^6(x^2 - 1)}{(x^2+1)^4 (x^2 + \beta^2)} \, dx
        \end{equation}
        By utilizing the algebraic identity $x^6 = x^4(x^2+\beta^2) - \beta^2 x^4$, we can elegantly reduce the dimensionality of $J(\beta)$ :
        \begin{align}
        J(\beta) &= \int_0^\infty \frac{x^4(x^2-1)}{(x^2+1)^4} \, dx - \beta^2 I_{\text{point}}(\beta) \\
        &= J(0) - \beta^2 I_{\text{point}}(\beta)
        \end{align}
        The constant integral $J(0)$ can be derived by the Cauchy Residue Theorem, which results in $J(0) = \frac{\pi}{8}$. Detailed derivation is provided in \ref{app:analytical_derivation}. The resonance peak occurs precisely when the net energy shift completely vanishes, which mathematically necessitates $I(\beta) = 0$. Equating our derived integrals :
        \begin{equation}
        I(\beta) - \gamma \left( \frac{\pi}{8} - \beta^2 I(\beta) \right) = 0 \implies I(\beta)(1 + \gamma \beta^2) = \gamma \frac{\pi}{8}
        \end{equation}
        Substituting the exact analytical form of $I(\beta)$ established earlier :
        \begin{equation}
        \frac{\pi \beta^2}{8(\beta+1)^4} (1 + \gamma \beta^2) = \gamma \frac{\pi}{8} \implies \beta^2 (1 + \gamma \beta^2) = \gamma (\beta+1)^4
        \end{equation}
        In the regime of the peak, the Yukawa range is far larger than the atomic scale ($\lambda \gg a_0$, hence $\beta \ll 1$), and the finite-size parameter can be considered to be small enough ($\gamma \sim \mathcal{O}(10^{-6})$). We can therefore confidently execute the leading-order truncation $(1 + \gamma\beta^2) \approx 1$ and $(\beta+1)^4 \approx 1$. The intricate cancellation condition reduces to an exceptionally elegant relation :
        \begin{equation}
        \beta^2 \approx \gamma
        \end{equation}
        Applying the definitions $\beta = a_0/\lambda$ and $\gamma = \frac{1}{6}(r_{nuc}/a_0)^2$, we obtain the analytical prediction for the interaction range where the resonance manifests :
        \begin{equation}
        \left( \frac{a_0}{\lambda} \right)^2 \approx \frac{1}{6} \left( \frac{r_{nuc}}{a_0} \right)^2 \quad \implies \quad \lambda \approx \sqrt{6} \frac{a_0^2}{r_{nuc}}
        \end{equation}
        For hydrogen and deuterium, the approximated Yukawa-range that resonance occurs is :
        \begin{itemize}
            \item \textbf{Hydrogen} : $\lambda \approx 2.37 \times 10^{5} \text{ fm}$
            \item \textbf{Deuterium} : $\lambda \approx 8.40 \times 10^{4} \text{ fm}$
        \end{itemize}
        While it is not exactly the same as what is derived from the numerical analysis due to the finite-size effect (since the form factor was approximated under the low-momentum limit), it provides a rigorous scaling analysis that is consistent with the observed outcome. At this critical range, the penalty induced by the finite-size effect perfectly cancels the residual Yukawa attraction difference between the $2s$ and $2p$ orbitals. Consequently, the Yukawa-type potential becomes "invisible" to the $2s-2p$ Lamb shift transition, producing no net observable energy displacement. This result forces the upper limit of the coupling constant $\alpha$ to diverge into the peaks observed in Fig.~\ref{fig:constraint_dip}.

\section{Conclusion}
In this work, we have established a highly precise analytical framework to constrain Yukawa-type fifth-force interactions using the $2s-2p$ Lamb shift in hydrogen and deuterium. By transforming the sixfold coordinate-space integral into a one-dimensional momentum-space integral, our approach significantly improves computational precision. Furthermore, this formulation is inherently model-independent, since the integration evaluates the form factors directly; our method can easily incorporate empirical data from elastic scattering experiments without relying on approximated nuclear charge distribution models.

The primary outcome of this analysis is the derivation of highly stringent constraints on the parameter space of the Yukawa coupling strength $\alpha$ and the interaction range $\lambda$. While collider and neutron scattering experiments are traditionally employed to constrain Yukawa-type potentials in the sub-femtometer range, we show that atomic spectroscopy provides an equally robust and independent probe.

Our results demonstrate that the transition frequency shifts induced by such new forces fall substantially below the resolution limits of current spectroscopic measurements. Consequently, the inherent precision of general spectroscopy constitutes the dominant bottleneck in probing these sub-femtometer interactions.

Within this constrained parameter space, we also observe a resonance-like cancellation effect between the $2s$ and $2p$ energy shifts at specific Yukawa ranges. This cancellation further suppresses the observable signal and manifests as a divergence in our constraint limits, mapping a clear boundary between attractive and repulsive interaction regimes. Beyond the resonance point, the Yukawa potential contributes to a repulsive interaction, while below the resonance point, it contributes to an attractive interaction.

Analysis of hydrogen and deuterium shows that the momentum-space framework is a robust tool in searches beyond the Standard Model. As experimental precision advances, this methodology translates empirical data into bounds on new physics. Ultimately, higher precision in both electronic and muonic systems will be needed to uncover or rule out fifth-force interactions in the attometer regime.

\appendix
\section{Analytical Derivation of the Integral for the Resonance-like Cancellation}
    \label{app:analytical_derivation}
    \subsection{The Integration for a Point-like Distribution}
        Since the integrand is an even function, we extend the integration to the entire real axis and evaluate it with the Cauchy Residue Theorem over the upper half-plane complex plane :
        \begin{equation}
            I_{\text{point}}(\beta) = \frac{1}{2} \int_{-\infty}^{\infty} \frac{z^4(z^2 - 1)}{(z^2+1)^4(z^2+\beta^2)} \, dz = \frac{1}{2} \oint_C f(z) \, dz
        \end{equation}

        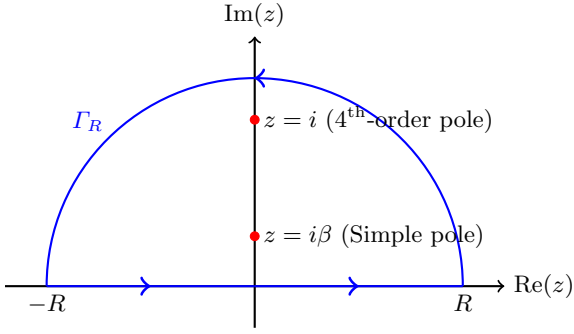
\begin{figure}[htbp]
            \centering
            \begin{tikzpicture}[scale=1.1, 
                decoration={markings, mark=at position 0.5 with {\arrow[scale=1.5]{>}}}]
                
                \draw[->, thick] (-3, 0) -- (3, 0) node[right] {$\text{Re}(z)$};
                \draw[->, thick] (0, -0.5) -- (0, 3) node[above] {$\text{Im}(z)$};
                
                \draw[thick, blue, postaction={decorate}] (-2.5, 0) -- (0, 0);
                \draw[thick, blue, postaction={decorate}] (0, 0) -- (2.5, 0);
                \node[below] at (2.5, 0) {$R$};
                \node[below] at (-2.5, 0) {$-R$};
                
                \draw[thick, blue, postaction={decorate}] (2.5,0) arc (0 :180 :2.5);
                \node[blue, above right] at (-2.3, 1.76) {$\Gamma_R$};
                
                \filldraw[red] (0, 0.6) circle (1.5pt) node[right, black] {$z = i\beta$ (Simple pole)};
                \filldraw[red] (0, 2.0) circle (1.5pt) node[right, black] {$z = i$ ($4^{\text{th}}$-order pole)};
                
            \end{tikzpicture}
            \caption{The complex contour $C$ for evaluating the integral $I(\beta)$. The contour consists of the real interval $[-R, R]$ and the semicircular arc $\Gamma_R$ in the upper half-plane. As $R \to \infty$, the integral over $\Gamma_R$ vanishes. The integrand has two poles, including a simple pole at $z=i\beta$ and a fourth-order pole at $z=i$.}
            \label{fig:contour}
        \end{figure}

        Two poles can be found, including a simple pole at $z = i\beta$ and a $4^{\text{th}}$-order pole at $z = i$. Direct evaluation of the residue for the fourth-order pole at $z=i$ is algebraically intractable. We introduce the variable transformation $w = z^2$ and apply partial fraction decomposition :
        \begin{equation}
            f(w) = \frac{w^2(w-1)}{(w+1)^4(w+\beta^2)} = \frac{A}{w+\beta^2} + \sum_{k=1}^4 \frac{B_k}{(w+1)^k}
        \end{equation}
        The coefficient $A$ is precisely determined by evaluating the residue at $w=-\beta^2$. The coefficient $A$ corresponding to the simple pole at $w = -\beta^2$ is isolated by multiplying the partial fraction expansion by $(w+\beta^2)$ and evaluating the residue :
        \begin{align}
            A &= \lim_{w \to -\beta^2} \left[f(w) \cdot (w+\beta^2) \right] \\
            &= \lim_{w \to -\beta^2} \frac{w^2(w-1)}{(w+1)^4}
        \end{align}
        Substituting $w = -\beta^2$ into the expression :
        \begin{equation}
            A = - \frac{\beta^4 (\beta^2 + 1)}{(1 - \beta^2)^4}
        \end{equation}
        This coefficient $A$ dictates the contribution of the simple pole at $z = \pm i\beta$ to the total integral.
        
        The coefficients $B_k$ are extracted by expanding the $f(w)$ in a Taylor series about $(w+1)$. To derive the coefficients $B_k$, we define a variable $v = w+1$. Substituting this into the function $f(w)$ excluding the $A$ term :
        \begin{equation}
            f(v) = \frac{(v-1)^2(v-2)}{v^4(v-\alpha)}
        \end{equation}
        where $\alpha = 1-\beta^2$. We isolate the numerator polynomial $P(v)$ and expand the denominator term $(v-\alpha)^{-1}$ as a geometric series. The polynomial is :
        \begin{equation}
            P(v) = (v-1)^2(v-2) = v^3 - 4v^2 + 5v - 2
        \end{equation}
        The geometric series expansion for $|v| < \alpha$ is given by :
        \begin{equation}
            \frac{1}{v-\alpha} = -\frac{1}{\alpha} \left( 1 + \frac{v}{\alpha} + \frac{v^2}{\alpha^2} + \frac{v^3}{\alpha^3} + \mathcal{O}(v^4) \right)
        \end{equation}
        We evaluate the combined non-singular numerator up to the third order in $v$ :
        \begin{equation}
            \frac{P(v)}{v - \alpha} = -\frac{1}{\alpha} \bigl( -2 + 5v - 4v^2 + v^3 \bigr)
            \bigl( 1 + \tfrac{v}{\alpha} + \tfrac{v^2}{\alpha^2} + \tfrac{v^3}{\alpha^3} + \dots \bigr)
        \end{equation}
        By expanding the polynomial into the infinite series and sort the resulting terms by their powers of $v$ (letting the result be expressed as $\sum c_n v^n$), we can extract each coefficient $c_n$ :
        \begin{align}
            c_0 &= \frac{2}{\alpha} \\
            c_1 &= \frac{2}{\alpha^2} - \frac{5}{\alpha} \\
            c_2 &= \frac{2}{\alpha^3} - \frac{5}{\alpha^2} + \frac{4}{\alpha} \\
            c_3 &= \frac{2}{\alpha^4} - \frac{5}{\alpha^3} + \frac{4}{\alpha^2} - \frac{1}{\alpha}
        \end{align}
        Since the original function is $f(v) = \frac{P(v)}{v^{4} (v - \alpha)}$, we divide this resulting Taylor expansion by $v^4$. This directly maps the Taylor coefficients $c_n$ to the partial fraction coefficients $B_k$ associated with $(w+1)^{-k}$. This yields the exact analytical matching :
        \begin{align}
            B_4 &= c_0 = \frac{2}{\alpha} \\
            B_3 &= c_1 = \frac{2}{\alpha^2} - \frac{5}{\alpha} \\
            B_2 &= c_2 = \frac{2}{\alpha^3} - \frac{5}{\alpha^2} + \frac{4}{\alpha} \\
            B_1 &= c_3 = \frac{2}{\alpha^4} - \frac{5}{\alpha^3} + \frac{4}{\alpha^2} - \frac{1}{\alpha}
        \end{align}
        The integration of these partial fraction terms along the complex contour is evaluated by the integral identity derived from the residue theorem for high-order poles :
        \begin{align}
            I_k &= \oint_C \frac{1}{(z^2+1)^k} \, dz \\
            &= 2\pi i \, \text{Res}\left( \frac{1}{(z^2+1)^k}, z=i \right) \\
            &= \frac{\pi}{2^{2k-2}} \binom{2k-2}{k-1}
        \end{align}
        We obtain the exact integral values :
        \begin{align}
            I_1 &= \pi, \quad I_2 = \frac{\pi}{2}, \quad I_3 = \frac{3\pi}{8}, \quad I_4 = \frac{5\pi}{16}
        \end{align}
        Let $L_B$ denote the total integral contribution from the $4^{\text{th}}$-order pole at $z=i$. Substituting the derived coefficients $B_k$ into the sum $L_B = \sum_{k=1}^4 B_k I_k$, we group the terms by the powers of $\alpha^{-1}$ :
        \begin{align}
            L_B &= B_1(\pi) + B_2\left(\frac{\pi}{2}\right) + B_3\left(\frac{3\pi}{8}\right) + B_4\left(\frac{5\pi}{16}\right) \nonumber \\
            &= \frac{\pi}{4\alpha^4} \left( 8 - 16\alpha + 9\alpha^2 - \alpha^3 \right)
        \end{align}
        Restoring the substitution $\alpha = 1-\beta^2$, the aggregate contribution from the pole at $z=i$ is :
        \begin{equation}
            L_B = \frac{\pi (\beta^6 + 6\beta^4 + \beta^2)}{4(1-\beta^2)^4}
        \end{equation}
        Similarly, the contour integral contribution $L_A$ from the simple pole at $z=i\beta$ which corresponds to the $A$ term can be evaluated directly via its residue :
        \begin{align}
            L_A &= \oint_C \frac{A}{z^2+\beta^2} \, dz \\
            &= 2\pi i \left( \frac{A}{2i\beta} \right) \\
            &= -\frac{\pi \beta^3 (\beta^2 + 1)}{(1-\beta^2)^4}
        \end{align}
        The total integral is the sum of the contributions of the poles, $L_{\text{total}} = L_A + L_B$ :
        \begin{align}
            L_{\text{total}} &= \frac{\pi}{4(1-\beta^2)^4} \Big[ (\beta^6 + 6\beta^4 + \beta^2) - 4\beta^3(\beta^2+1) \Big] \nonumber \\
            &= \frac{\pi \beta^2}{4(\beta+1)^4}
        \end{align}
        Finally, since the integrand is an even function, the target integral $I_{\text{point}}(\beta)$ from $0$ to $\infty$ is exactly half of $L_{\text{total}}$ :
        \begin{equation}
            I_{\text{point}}(\beta) = \frac{1}{2} L_{\text{total}} = \frac{\pi \beta^2}{8(\beta+1)^4}
        \end{equation}
    \subsection{The Constant Integral for a Finite-size Distribution}
        We present a rigorous step-by-step derivation of the integral
        \begin{equation}
            J(0) = \int_{0}^{\infty} \frac{x^4(x^2 - 1)}{(x^2 + 1)^4} dx.
        \end{equation}
        Since the integrand $f(x) = \frac{x^4(x^2 - 1)}{(x^2 + 1)^4}$ is an even function, we can extend the integration domain to the entire real line :
        \begin{equation}
            J(0) = \frac{1}{2} \int_{-\infty}^{\infty} \frac{z^4(z^2 - 1)}{(z^2 + 1)^4} dz.
        \end{equation}
        We extend the real variable $x$ to the complex variable $z$ and consider the closed semicircular contour $C$ in the upper half-plane, which consists of the real axis from $-R$ to $R$ and a semicircular arc $\Gamma_R$ of radius $R$. By the Residue Theorem :
        \begin{equation}
            \oint_C f(z) dz = 2\pi i \sum_{k} \text{Res}(f, z_k).
        \end{equation}
        The function $f(z)$ has $4^{\text{th}}$-order poles at $z = \pm i$. Within our chosen contour in the upper half-plane, the only enclosed singularity is a pole at $z = i$. For a pole of order $n$, the residue is calculated as :
        \begin{equation}
            \text{Res}(f, i) = \frac{1}{(n-1)!} \lim_{z \to i} \frac{d^{n-1}}{dz^{n-1}} \left[ (z-i)^n f(z) \right].
        \end{equation}
        Substituting $n=4$ :
        \begin{equation}
            \text{Res}(f, i) = \frac{1}{3!} \lim_{z \to i} \frac{d^3}{dz^3} \left[ \frac{z^4(z^2-1)}{(z+i)^4} \right].
        \end{equation}
        Let $g(z) = \frac{z^6-z^4}{(z+i)^4}$. The residue is exactly the coefficient of the third-order term in the Taylor series expansion of $g(z)$ around $z=i$. We introduce the change of variable $w = z - i$, so the expansion point becomes $w = 0$ :
        \begin{equation}
            g(w+i) = \frac{(w+i)^6 - (w+i)^4}{(w+2i)^4}.
        \end{equation}
        We expand the numerator $N(w)$ and the denominator $D(w)$ up to $\mathcal{O}(w^3)$. For the numerator :
        \begin{equation}
            N(w) = -2 + 10iw + 21w^2 - 24iw^3 + \mathcal{O}(w^4).
        \end{equation}
        For the denominator :
        \begin{equation}
            D(w)^{-1} = \frac{1}{16} \left( 1 + 2iw - \frac{5}{2}w^2 - \frac{5i}{2}w^3 + \mathcal{O}(w^4) \right).
        \end{equation}
        Multiplying the respective terms that sum to order 3, we get :
        \begin{equation}
            c_3 = -\frac{i}{8}.
        \end{equation}
        Therefore, $\text{Res}(f, i) = -i/8$. Applying the Residue Theorem, the contour integral yields :
        \begin{equation}
            \oint_C f(z) dz = 2\pi i \left( -\frac{i}{8} \right) = \frac{\pi}{4}.
        \end{equation}
        Substituting this back into our initial expression for $J(0)$, we obtain the final result :
        \begin{equation}
            J(0) = \frac{1}{2} \left( \frac{\pi}{4} \right) = \frac{\pi}{8}.
        \end{equation}

\bibliographystyle{unsrt}
\bibliography{reference}

\end{document}